# A Latency-Aware Real-Time Video Surveillance Demo: Network Slicing for Improving Public Safety

**B. Shariati[1,*], J. J. Pedreno-Manresa[2], A. Dochhan[2], A. S. Muqaddas[3], R. Casellas[4], O. González de Dios[5], L. L. Canto[5], B. Lent[6], J. E. López de Vergara[7], S. López-Buedo[7], F. J. Moreno[8], P. Pavón[8], L. Velasco[9], S. Patri[2], A. Giorgetti[10], F. Cugini[10], A. Sgambelluri[10], R. Nejabati[3], D. Simeonidou[3], R,-P, Braun[11], A. Autenrieth[2], J.-P. Elbers[2], J. K. Fischer[1], R. Freund[1]**

*1 Fraunhofer HHI, Germany; 2 ADVA, Germany; 3 UOB, UK; 4 CTTC/CERCA, Spain; 5 Telefónica, Spain; 6 Qognify GmbH, Germany; 7 Naudit HPCN, Spain; 8 UPCT, Spain; 9 UPC, Spain; 10 CNIT, Italy; 11 Deutsche Telekom, Germany*

[*] behnam.shariati@hhi.fraunhofer.de

**Abstract:** We report the automated deployment of 5G services across a latency-aware, semi-disaggregated, and virtualized metro network. We summarize the key findings in a detailed analysis of end-to-end latency, service setup time, and soft-failure detection time.

## 1. Introduction

One of the promises of 5G networks is to deliver low-latency end-to-end services in a very short time upon the customer's request [1]. While considering a metro-scale network, such a goal necessitates a sophisticated optical networking solution able to dynamically instantiate end-to-end network slices, where sufficient resources (i.e., connectivity and computing) are assigned to the incoming request with minimum service setup time. Moreover, the operation of such a network should be monitored to anticipate failures, thus avoiding any service disruption [2].

We designed and built a smart optical metro infrastructure able to deliver requirements for time-critical and high-bandwidth vertical use cases, including public safety ones [3]. The development comprises several hardware and software pieces that together form a unique networking infrastructure, including both packet and optical layers, along with an integrated NFV and disaggregated network orchestration, which provides advanced connectivity services and ETSI Network Services (NS) encompassing computing, storage and networking resources [4].

This paper adapts the control and management architecture presented in [4], extends it with several key features, including end-to-end latency-awareness as well as monitoring and data analytics capabilities, to operate a partially disaggregated edge computing enabled metro optical network. We then use the developed infrastructure to demonstrate, in a lab and in a field trial, a real-time and latency-aware video surveillance vertical use-case that benefits from autonomous deployment of 5G network services. The considered video surveillance use-case requires high-bandwidth connectivity for streaming video footage, edge computing resources to run its on-the-edge video analytics micro services, and very low-latency communication to allow real-time feedback to the cameras, upon any decision from the analytics engine. The architecture enables autonomous object tracking and intrusion detection using on-camera and on-the-edge analytics, respectively. More details of the demonstration scenarios are available in [3]. The results show that the end-to-end latency over an 80 km link and service deployment time are less than 800 µs and 180 seconds, respectively, meeting the Key Performance Indicators (KPI) defined by 5G PPP.

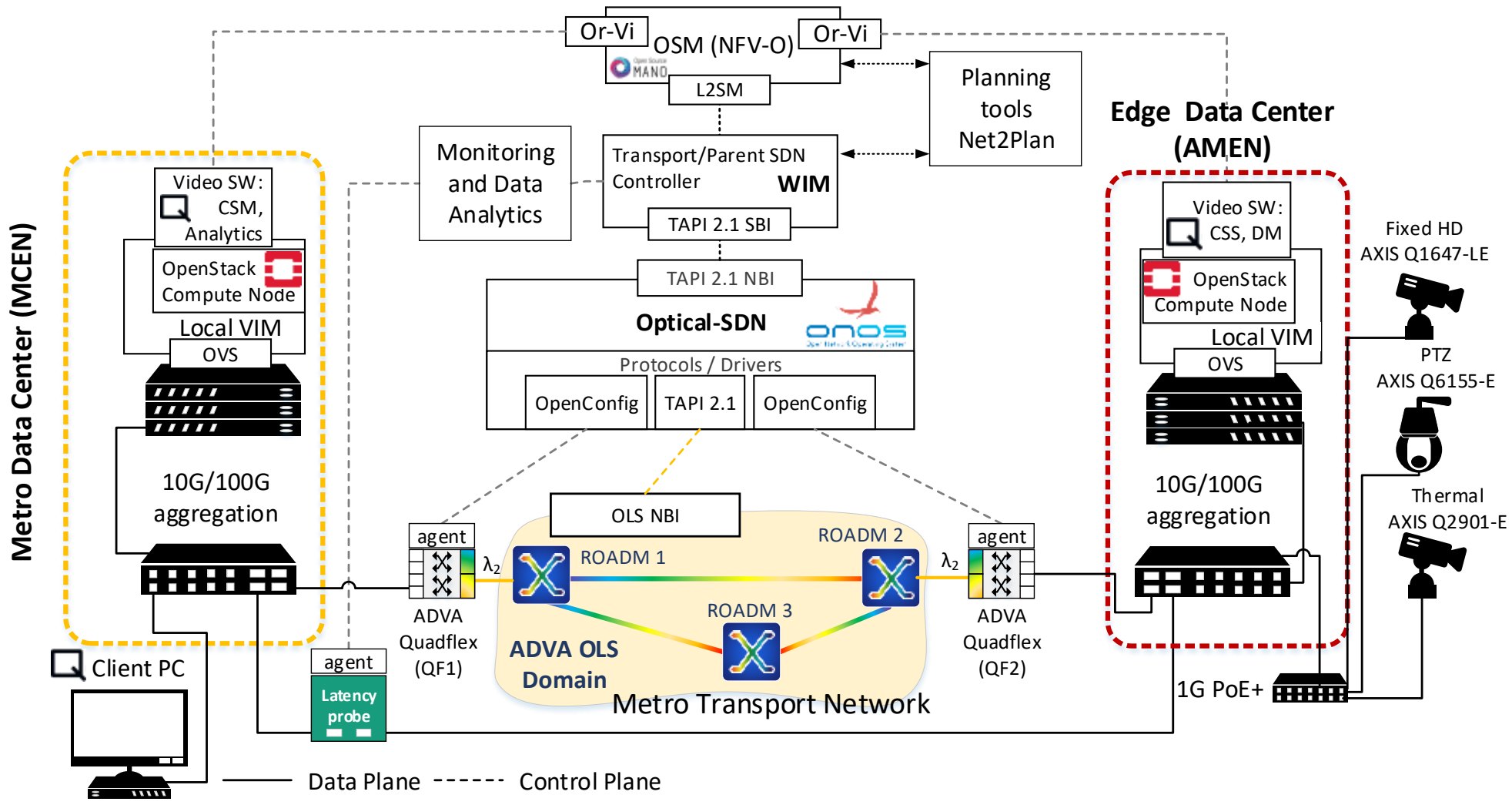

**Fig. 1:** Architecture of the Demonstration

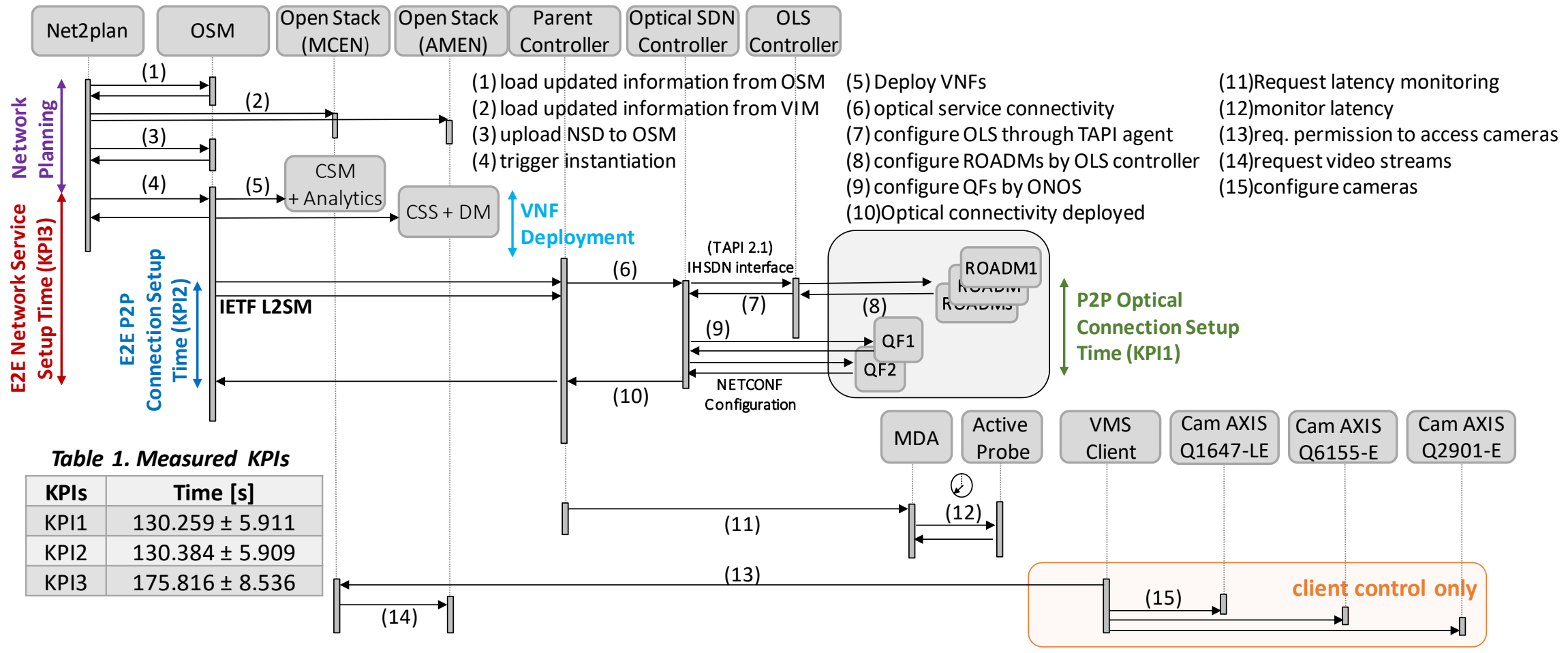

*Table 1. Measured KPIs*

| KPIs | Time [s] |
|---|---|
| KPI1 | 130.259 ± 5.911 |
| KPI2 | 130.384 ± 5.909 |
| KPI3 | 175.816 ± 8.536 |

**Fig. 2:** Workflow of the Demonstration

## 2. Demonstration Architecture and Scenario

Fig.1 shows the demonstration architecture. Our proposed architecture adapts parts of the control plane components presented in [4] to a new and different data plane setup. We setup a partially disaggregated optical network comprising three commercial 2-degree semi-filterless reconfigurable optical add-drop multiplexers (ROADM), based on wavelength blockers and splitters [5], interconnected in a ring topology and two commercial coherent transponders [6]. Each one of the transponders is connected to a compute node via an aggregation switch; each compute node acts as an edge datacenter.

From a control plane perspective, we benefit from Net2Plan [7], Open Source MANO (OSM), OpenStack, parent SDN controller, and optical SDN controller modules as presented in [4]. However, we extend it in several key aspects. In this work, we follow a partially disaggregated scenario, which requires a hierarchy of optical SDN controllers. Our proposed optical SDN controller (i.e., ONOS) manages the ROADMs through the pen Line System (OLS) controller using TAPI South Bound Interfaces (SBI), while it manages the transponders using OpenConfig SBI. For both SBIs a specific driver has been developed. Moreover, our demonstration incorporates a Monitoring and Data Analytics (MDA) [8] controller that carries out fault degradation analysis and latency measurement retrieval upon a request from the parent SDN controller. Another key novelty of our setup is the latency measurement capability, which becomes possible by an active 100G probe [9] that provides real-time round-trip latency measurements. Our infrastructure also includes three cameras and their corresponding distributed Video Management System (VMS), which together realize a real-time video surveillance use-case.

Fig. 2 shows the demonstration workflow, which shares some similarities with the one presented in [4]. The workflow has been extended to highlight the interactions of the additional modules. The additional interactions include 1) the ones between ONOS and OLS controller to control the ROADMs, 2) the ones among the active probe, the MDA controller, and the parent SDN controller, and 3) the ones among the cameras in the surveillance zone and the corresponding VNFs instantiated by OpenStack. One of the VNFs hosts the Core System Master (CSM) and the video analytics of the VMS, while the other one hosts the Core System Slave (CSS) and the Device Manager (DM)).

## 3. Results and Discussion

To assess the performance of the system, we report some of the key results obtained during the demonstration. From the data plane perspective, the optical transponders multiplex the client traffic into a 100 Gb/s DP-QPSK signal. The signal is then transmitted from ROADM1 to ROADM2 directly. The link length between ROADM1 and ROADM2 was varied between 6.8 km field-deployed fiber of the Deutsche Telekom R&D SASER TestNet, 41 km, and 80 km, for measuring end-to-end latency. In order to test probing of minimal latency values and calibrate measurements, also a 2.1 m long fiber was considered. Next, we present the evaluated service setup time, QoS metrics, and soft-failure degradation time.

**Service Setup Time**: we report three different KPIs as shown and defined in Fig. 3. We repeated the experiments 28 times to obtain a statistically reliable value for each KPI, which characterize the service deployment time. The definition of each KPI is provided in Fig. 3. The measured KPIs are listed in the Tab. 1 inset of Fig. 2, while a detailed version of different phases is illustrated in Fig. 3. Note that, while the configuration of the optical transponders is performed in ~2 s, their transmitting laser takes a significant time (~125 s) to warm up and stabilize. Excluding the transponders, the end-to-end service setup time takes ~50 s.

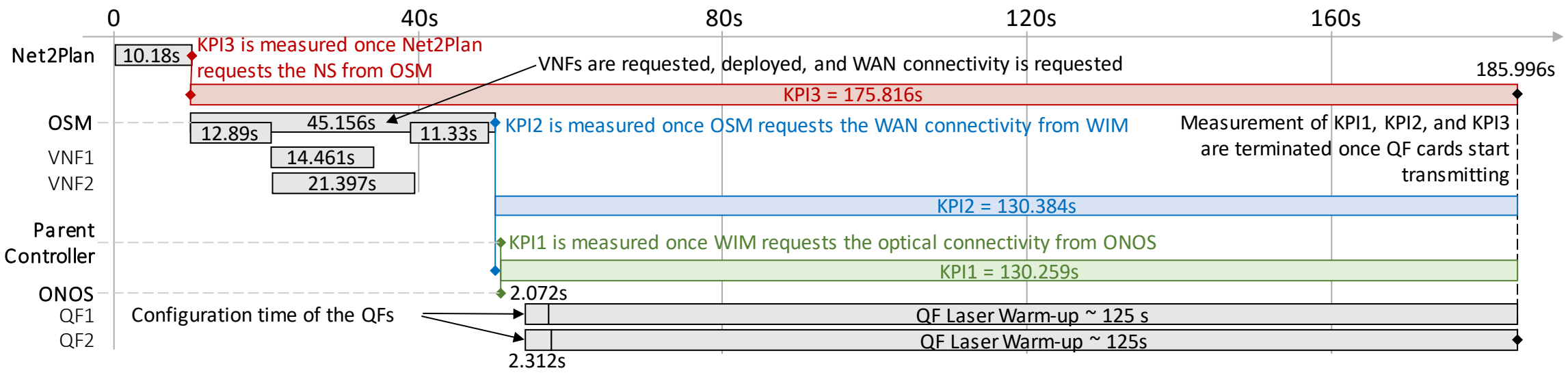

**Fig. 3:** Service setup time of different stages of the demonstration

**Table 2:** End-to-end Latency Measurements

| Link Length [km] | Measured Latency [μs] | Estimated Round Trip Propagation Delay [μs] | Delta [μs] |
|---|---|---|---|
| 0.0021 | 15.257 | 0.021 | 15.236 |
| 41.3665 | 420.453 | 405.044 | 15.409 |
| 79.9695 | 799.188 | 783.158 | 16.030 |
| 6.8 | 799.202 | 66.595 | 732.607 |

**Table 3:** Fault Degradation Time

| | Case 1 | Case 2 |
|---|---|---|
| Detection time | 4.46 min | 1.06 min |
| Anticipation time | 12 min | 1.06 min |
| Mean detection SNR | 15.15 dB | 14.28 dB |
| Mean detection BER | $1.36\ 10^{-8}$ | $2.97\ 10^{-7}$ |

**End-to-end Latency Measurement:** Tab. 2 summarizes the latency analysis. The 2.1 m link provides a way to calibrate our measurements. The latency of 41 km and 80 km links are dominated by the propagation time. Based on the delta of the three first measurements shown in Tab. 2, we derive the latency budget of the different elements of our test-bed: 0.84 μs is added by the probe, 1.29 μs by the aggregation switches, and 13.1 μs by the optical path devices. Note that, the delta of the field-deployed fibre is quite large, which is due to the many intermediate old fiber patches and extra devices on the fibre link interconnecting our test-bed to the Deutsche Telekom network in Berlin.

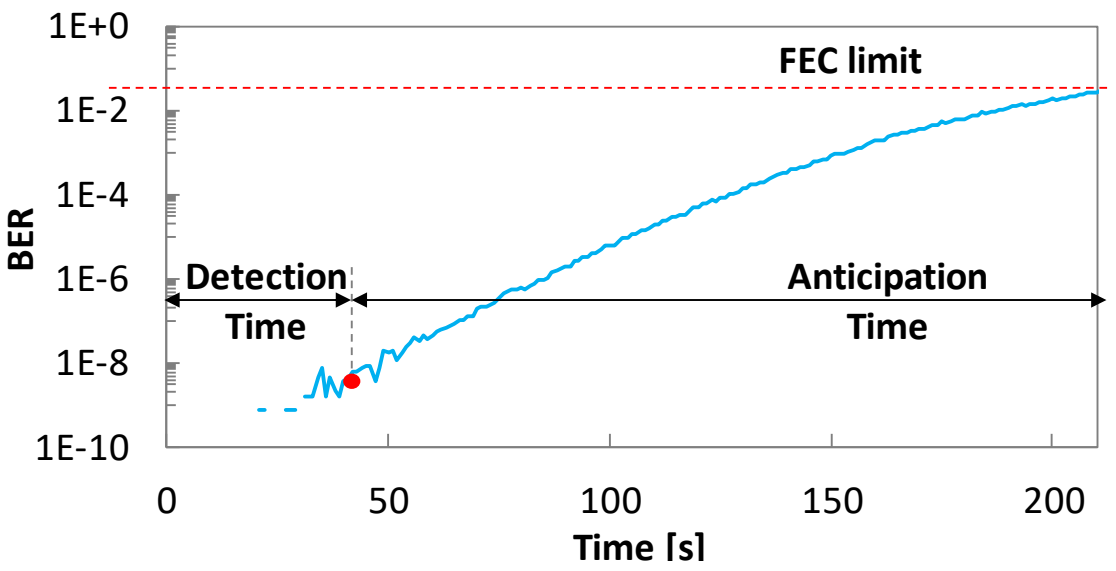

**Fig. 4:** BER degradation of one measurement of case 1

**Soft-failure Degradation Time:** Here, the intention is to anticipate QoT degradation, leaving enough time to implement counter-measures. We introduced intentional fibre loss by an additional variable optical attenuator to emulate the failure, first by 0.025dB per second (case 1) and then by 0.25dB per second (case 2).The effects have been measured in terms of SNR and pre-FEC BER at the receiver. The BER degradation of a single measurement is illustrated in Fig. 4, while Tab. 3 presents the average value of all the measurements. The anticipation time after detection leaves more than one minute to reconfigure the degraded optical connection.

## 4. Conclusion

We successfully demonstrated a video surveillance vertical use-case in a latency-aware metro network with end-to-end service setup and network slicing including connectivity and virtual network functions. A multi-partner, multi-component integration based on software and hardware elements, including Net2plan, OSM, ONOS, and OpenConfig enabled open terminals was achieved. We have demonstrated the integration of an NFV MANO architecture with a hierarchical, multilayer SDN control plane for disaggregated networks, including a dedicated OLS controller. We have evaluated the performance of the system, both in a lab trail and in a field trial, including not only the dynamic provisioning but also latency-awareness as well as subsequent monitoring and fault detection processes. Finally, KPIs have been experimentally quantified highlighting the key benefits and asserting the key role of the optical technology for services with stringent latency and bandwidth requirements.

**Acknowledgments:** The research leading to these results has received funding from the EC and BMBF through the METRO-HAUL project (G.A. No. 761727) and OTB-5G+ project (reference No. 16KIS0979K).

## References

[1] 5G PPP Architecture Working Group, "View on 5G Architecture," version 3.0, Feb 2020 (Available https://5g-ppp.eu/white-papers/).
[2] A. P. Vela, et al, "Soft failure localization during commissioning testing and lightpath operation," JOCN, 10(1), 2018.
[3] A. Dochhan, et al., "Metro-haul project vertical service demo: video surveillance real-time low-latency object tracking," OFC, 2020.
[4] A. S. Muqaddas, et al., "Field trial of multi-layer slicing over disaggregated optical networks enabling end-to-end crowdsourced video streaming," ECOC, 2020.
[5] ADVA FSP3000 Open Line System with microROADMs (https://www.adva.com/) [accessed in Aug 2020]
[6] ADVA Quadflex Transponder (https://www.adva.com/) [accessed in Aug 2020]
[7] Net2plan, "The Open Source Planner," [Online]. Available: www.net2plan.com. [Accessed Nov 2020]
[8] L. Velasco, et al., "Monitoring and data analytics for optical networking: benefits, architectures, and use cases," IEEE Network, 2019.
[9] Naudit Active Probe [Online]. Available: http://www.naudit.es/en/ [Accessed in Nov 2020]